\documentclass{llncs}
\pdfoutput=1

\usepackage{llncsdoc}
\usepackage{lineno,hyperref}
\modulolinenumbers[5]
\usepackage{mathptmx}       
\usepackage{helvet}         
\usepackage{courier}        
\usepackage{makeidx}         
\usepackage{multicol}        
\usepackage[bottom]{footmisc}
\usepackage{color}

\usepackage{times,amsmath,epsfig}
\usepackage{graphicx}
\usepackage[T1]{fontenc}
\usepackage{algorithmic}

\usepackage[vlined,linesnumbered,ruled]{algorithm2e}
\usepackage{multirow}
\usepackage{pdfpages}
\usepackage{epstopdf}
\usepackage{amsmath}
\usepackage{amssymb}
\usepackage{subfigure}
\usepackage{slashbox}

\usepackage{times,amsmath,epsfig}

\begin{document}

\title{A Robust Process to Identify Pivots inside Sub-communities In Social Networks}
\author{Joseph Ndong\inst{1} \and Ibrahima Gueye\inst{2}}
\institute{Department of Mathematics and Computer Science\\University Cheikh Anta Diop, BP 5005 Fann Dakar, Senegal
\and
 Computer Engineering and Telecom, Polytechnic School, Thies, Senegal \\joseph.ndong@ucad.edu.sn, igueye@ept.sn}

\maketitle

\begin{abstract}
In this work, we extend a previous work where we proposed a suitable state model built from a Karhunen-Loeve Transformation to build a new decision process from which, we can extract useful knowledge and information about the identified underlying sub-communities from an initial network. The aim of the method is to build a framework for a multi-level knowledge retrieval. Besides the capacity of the methodology to reduce the high dimensionality of the data, the new detection scheme is able to extract, from the sub-communities, the dense sub-groups with the definition and formulation of new quantities related to the notions of energy and co-energy. The energy of a node is defined as the rate of its participation to the set of activities while the notion of co-energy defines the rate of interaction/link between two nodes. These two important features are used to make each link weighted and bounded, so that we are able to perform a thorough refinement of the sub-community discovery. This study allows to perform a multi-level analysis by extracting information either per-link or per-intra-sub-community. As an improvement of this work, we define the notion of pivot to relate the node(s) with the greatest influence in the network. We propose  the use of a thorough tool based on the formulation of the transformation of a suitable probabilistic model into a possibilistic model to extract these pivot(s) which are the nodes that control the evolution of the community.
\end{abstract}


%

\section{Introduction}
\begin{figure}[!h]
\centering
\includegraphics[height=120pt]{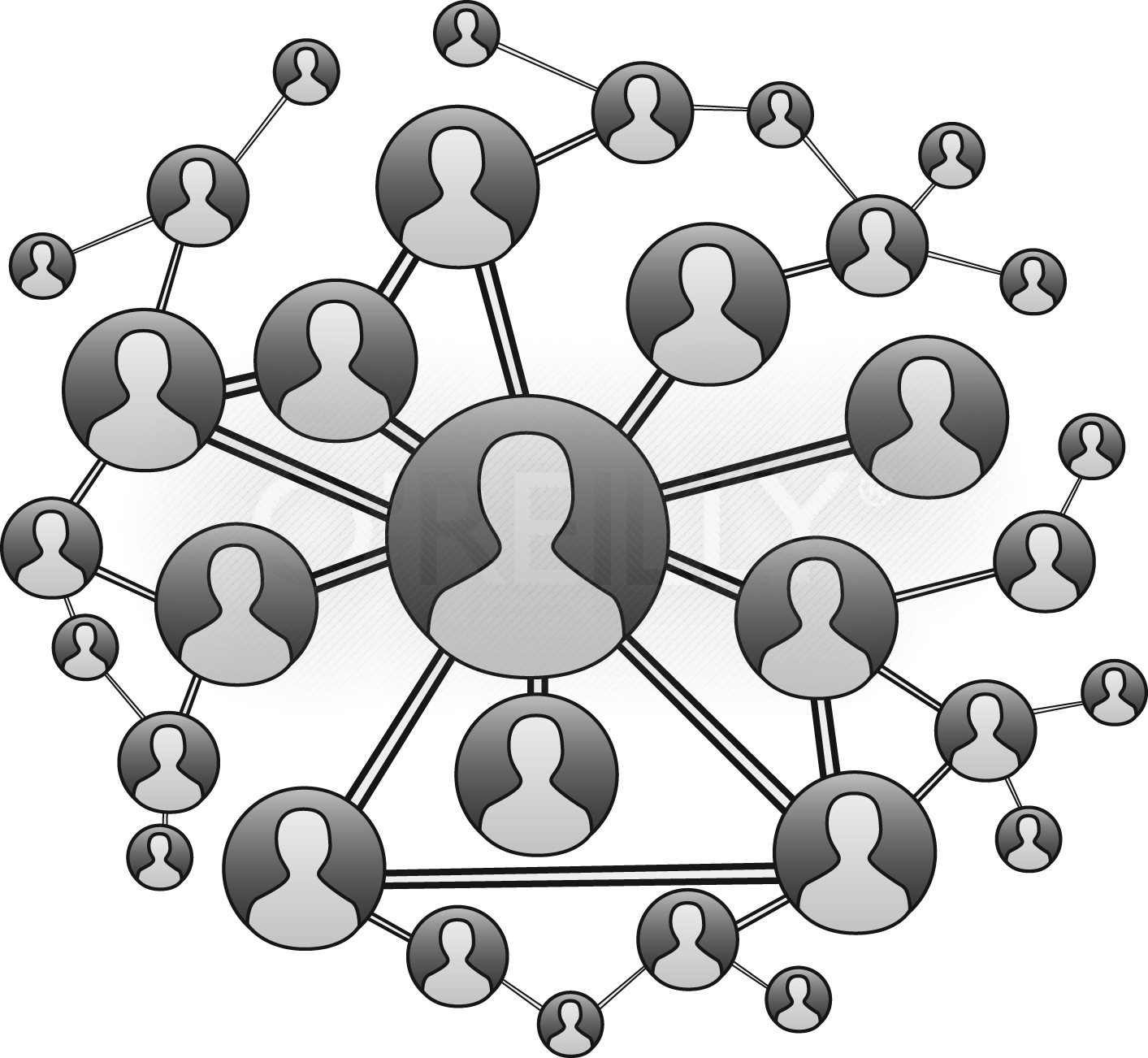}
\caption{Social Network Showing Nodes and Links (Source: \cite{Sorour2014})}
\label{snf}
\end{figure}

Social networks describe web-based services that allow  users (individuals) to connect with other users, communicate, share or publish contents within the network \cite{AdedoyinOlowe2014}, \cite{chenZS}. The rise of web 2.0 has come with the ease in the production and sharing of content that allowed the social networks development. Users themselves become the producers of web content \cite{AdedoyinOlowe2014}, \cite{Kaplan}. In a formal and simple description, social network can be seen as a graph consisting of nodes (as individuals) and links (as social links) used to represent social relations on social network sites \cite{BorgattiSP} (as presented in Figure \ref{snf}).

These last years, Social networks sites have become some important sources of online interactions, contents sharing \cite{ChelmisC} and communication means. While giving these opportunities, they are affordable and universally acclaimed.  
Social network sites are commonly known for information dissemination, personal activities posting \cite{AsurS}, product reviews, online pictures sharing, professional profiling, advertisements, subjectivity \cite{AsurS}, assessments \cite{KimY}, approaches \cite{Korda}, influences \cite{BakshyE}, observations \cite{ChouW}, feelings \cite{Kaplan}, opinions and sentiments expressions \cite{PangB}, news, remarks, reactions, or some many other content type \cite{LiuB}. News alerts, breaking news, political debates and government policy are also posted and analyzed on social network sites.\\
Obervations \cite{Statista} show that more and more people are becoming interested and involved in social networks. Sometimes, users of social networks exploit them for making decisons. These decisions can be maked based on know or famialiar users or, in some case, based on unfamiliar users \cite{PangB}. 
Such a situation increases the degree of confidence in the credibility of these sites. The social network has transformed the way different entities procure and retrieve valuable information regardless of their location. The social network has also given users the privilege of giving their opinion.

The massive data generated or produced by all the social networks users allows for thorough analysis to efficiently extract useful knowledge such as trends, opinion leaders, influencers, feelings, etc.  Such an analysis can be done as a whole, which means the representations of all the social network actors then proceed to the detection of communities and/or opinion leaders. The main aim of community detection  methods is to partition the network into dense regions of the graph. Those dense regions typically correspond to entities which are closely related. The determination of such communities is useful in the context of a variety of applications such as customer segmentation, recommendations, link inference and influence analysis \cite{Khatoon2015}. 

In this work, we focus on community  influencers identification. We extend our previous work \cite{ndong2016new} in which, we derive a suitable state model from Karhunen-Loeve Transformation \cite{gray2004intro}. From this state model we built a new decision process, which allows us to extract useful knowledge and information about the identified underlying sub-communities from an initial network. 

This extension is interested in the extraction of the most "important" nodes in given sub-communities already detected or known. 
Identifying sub-groups under a social network is a great challenge in the area of social network analysis \cite{Becker2011}, \cite{Fortunato2010}, \cite{sarrjoe2014overlaying}, but analyzing the intrinsic behavior of each sub-group might be an important need if one desires to focus on the quality of nodes/actors \cite{AgrawalC}. If sub-communities are identified, decisions could be taken on a group of nodes regardless the underlying contents of the group itself since it can be viewed as a single homogeneous entity in which all items are of same behavior. One could also see the sub-community as a machine where the dynamic structure and the evolution depend strongly on the nature and quality of its combined pieces. If this second point-of-view is adopted, we need a tool to identify the "most important" pieces whose actions are more influential than those of the rest of the group. We define these pieces as "pivots" and the quality of the network is as important as it contains influential pivots with high degree of influence. 
In the following, we define pivots and how they could be important for a manager, a advertiser and so on.  

Inside a sub-community, nodes share links so, identifying pivots can be helpful in many situations:  \begin{itemize}
\item (i) if in a sub-community, the pivot(s) is/are linked to a few other nodes, the manager can conclude that this network is of less quality simply because the "most" important nodes have less interactivity and thus less influence. But in the contrary, he can pay more attention to the underlying evolution of the group; \\
\item (ii) the different sub-communities can be classified in an ascending order based on their quality --  the first group is the one with the greater number of pivots, and so on; \\
\item (iii) in a sub-community, we can find other sub-communities in the following manner: If the pivots themselves are not linked to each other, each of them can be the center of a group/entity formed by all the nodes linked to a pivot.
\end{itemize}

The procedure for sub-communities detection/identification we built in our previous work \cite{ndong2016new} consisted in extracting a decision variable from a state-space model we defined with a Karuhen-Loeve Transformation (KLT).  This have served also to reducing the dimensionality of the dataset in order to maintain the only relevant part of the data. Then, after defining some new quantities as "energy" of an actor/node and the "co-energy between" actors, we apply a decision process to identify the sub-communities and the features inside each of them. 
The improvement we bring in this present work is the ability to learn more about the impact of node's energy to see if some  nodes could be selected as those with the highest degree of influence. The notion of highest degree must be properly define to achieve this aim. We found that a suitable definition of a probability density function (pdf) related to the defined node's energy can be done. Then,  a second definition of a possibilistic model will be able to discover the relevant node(s) we classify as pivot(s).

Thes rest of this paper is organized as follow: In section \ref{background} we do a thorough study of the related work; in section \ref{method} we detail the methodology and aogorithm for the sub-commumity detection; in section \ref{possibility} we present our use of possibility theory to identifiy the pivots; in Section \ref{validation} we present the results of our experimental validation of our propositions, and we conclude this paper in Seciton \ref{conclusion}.

\section{Background}
\label{background}
Social sites have undoubtedly bestowed unimaginable privilege on their users to access readily available never-ending uncensored information. Twitter, for example, permits its users to post events in real time way ahead the broadcast of such events on traditional news media. Also, social network allow users to express their views, be it positive or negative \cite{AgrawalC}. Organizations are now conscious of the significance of consumers' opinions posted on social network sites to the patronage of their products or services and the overall success of their organisations. On the other hand, important personalities such as celebrities and government officials are being conscious of how they are perceived on social network. These entities follow the activities on social network to keep abreast with how their audience reacts to issues that concerns them \cite{CastellanosM}. 

Opinion of influencers on social network is based largely on their personal views and cannot be hold as absolute fact. However, their opinions are capable of affecting the decisions of other users on diverse subject matters. For example, Rihanna's rant has cost Snapchat \$800 million in one Day \cite{ririvssnap}. Opinions of influential users on Social network often count, resulting in opinion formation evolvement. Clustering technique of data mining can be used to model opinion formation by  assessing the affected nodes and unaffected nodes. Users that depict the same opinion are linked under the same nodes and those with opposing opinion are linked in other nodes. This concept is referred to as homophily in social network \cite{McPhersonM}. Homophily can also be demonstrated using other criteria such as race and gender \cite{Jackson2008}.

Researchers in social network analysis are facing many other research issues and challenges such as in \textit{Linkage-based and Structural Analysis}, or in \textit{Dynamic Analysis and Static Analysis}.

Linkage-based and Structural Analysis consists of anlyzing of the linkage behaviour of the social network so as to ascertain relevant nodes, links, communities and imminent areas of the network - Aggarwal (2011) \cite{AgrawalC}. 

Static analysis, such as in bibliographic networks, is presumed to be easier to carry out than those in streaming networks. In static analysis, it is presumed that social network changes gradually over time and analysis on the entire network can be done in batch mode. Conversely, dynamic analysis of streaming networks like Facebook and YouTube are very difficult to carry out. Data on these networks are generated at high speed and capacity. Dynamic analysis of these networks are often in the area of interactions between entities such as in Papadopoulos et al. 2012  \cite{Papadopoulos2012} and Sarr et al. 2014 \cite{Sarr2014}. Dynamic analysis is also covered through temporal events on social networks such as in Adedoyin-Olowe et al (2013) \cite{AdedoyinOlowe2013} and Becker et al (2011) \cite{Becker2011}, and we also have such dynalysis through the study of evolving communities - Fortunato, (2010) \cite{Fortunato2010}, Sarr et al. 2014 \cite{sarrjoe2014overlaying}.

\section{Methodology and algorithm for the sub-communities detection}
\label{method}
This methodology tracks and detects sub-communities based on the analysis of a huge number of features corresponding to events/activities for which a group of actors/nodes participate. First, we aim at finding the main features, to incorporate in our model, by means of extended principal component analysis. The second relevant issue of this methodology is related to the specification of a new detection procedure consisting of merging all the relevant features into a single process we will label as a "Decision Variable" (DV). By analyzing this process for the sub-community tracking operation, we can discover subgroups of actors using a multi-level thresholding and the notion of "energy  dissipation" of an actor over the events.

We consider a community of $R$ actors $\Omega=(a_1,\ldots,a_R)$ which perform activities on a set of $K$ initial correlated events $(e_1,\ldots,e_K)$. For each event $e_k$, we have a column vector of size $R$ containing the amount of participation of all $R$ actors to the corresponding activity. This operation gives us the $R\times K$ matrix of correlated random variables $X = (X_1, \ldots , X_K)$. In other words, one observes these random variables through $R$ independent realization vectors 
$x^i = (x_1^i,\ldots,x_K^i)$ $i = 1,\ldots, R$.\\
After extracting the relevant components from the Karhunen-Loeve transformation \cite{ndong2016new}, we can build our decision variable as a  row vector $DV=(y_1,\ldots,y_K)$. Then we can set a certain number of concepts for our methodology. We introduce the notion of "energy dissipation" (Ed) to quantify the degree of importance a given actor puts on a series of events. This notion is simple and intuitive. When considering the set of events/activities, the events for which the actor puts a high degree of importance constitutes his energy. For example, we can consider money as energy. When someone goes to buy some products, we can say that he/she is dissipating a certain amount of his/her energy.  In this case, he/she should buy a "product A", and consequently buy another "product B" necessary to use the product A. Here, we can see the notion of correlation between these products/variables.  When an athlete performs several disciplinary exercises in sport, we can view his actions as the dissipation of his energy over the different events, in order to win a medal. The energy of an actor is then quantifiable, its a measure of the strength of his participation to the a series of activities. 

If the actor participates actively to all or most of the activities with a high intensity, then his energy increases, otherwise we say that this actor has less energy according to the ensemble of events happening at a given period of time. 

Since the DV variable contains the aggregated amount of all actors participation to all events,  the energy dissipation $Ed$ of an actor $i$ is the row vector defined as:
\begin{equation}
\label{energydissip}
Ed_{i} = \left\lbrace  k,/ x_{k}^{i} \geq \text{DV}[k], \forall k=1,\ldots,K \right\rbrace  
\end{equation}

$Ed_i$ contains all the index of events for which the energy dissipation is greater than the reference DV.
Consequently, we can calculate the total energy of the actor $i$ as:
\begin{equation}
\label{energyactor}
E_{i}=\frac{|Ed_{{i}}|}{|DV|}
\end{equation}
where $|.|$ indicates the size of a vector. 

We also refer to the notion of "co-energy" dissipation (CED) as the amount of energy between two actors according to their participation to the same set of activities.  This quantity is a measure of the mean  energy produced simultaneously by the two actors on the same activities: 
\begin{equation}
\label{coenergydissip}
\text{CED}_{ij}=\frac{|\left(Ed_{i}\cap Ed_{j}\right) |}{|DV|}
\end{equation}

Finally, our detection procedure boils down to fixe a threshold $\alpha$ and put a link between actor $i$ and actor $j$ if the rate of their co-energy exceeds  the limit $\alpha$. This means the following inequality must be held to add the link:
\begin{equation}
\label{linkactor}
\text{CED}_{ij}\geq \alpha
\end{equation}

When Eq. \ref{linkactor} holds, the value of $\text{CED}_{ij}$ becomes the weight of the link between actor $i$ and actor $j$. And then, this link is bounded by the interval $\left[min(E_i,E_j),max (E_i,E_j)\right] $.
By varying the threshold $\alpha \in [0;1]$ , one can build many different sub-communities with the same dataset, each sub-community with a score $\alpha$ which measures its degree of realization. The algorithm to achieve our aim is described as follow:
 \begin{algorithm}
     \SetKwInOut{Data}{Parameters}
     \SetKwInOut{Input}{Input}
     \SetKwInOut{Output}{Output}
    \footnotesize{
     \Input{$C_t$, a community\\$\Omega(C_t)$, the sets of actors within $C_t$\\
      $x^i = (x_1^i,\ldots,x_K^i)$ the vector of participation of actor $i$ \\
     $DV=(dv_1,\ldots,dv_K)$ the decision variable\\$\alpha$,  the link detection threshold}
     \Output{\textbf{V}, a sub-community}
     \CommentSty{ /* Calculate Co-Energy dissipation between actors and apply threshold to add link*/} \\
     \Begin{
     \ForEach { $(k, l) \in \Omega(C_t)$, $k\neq l$}
		{      \CommentSty{ /* Apply Eq. (\ref{energydissip})*/} \\
		    $Ed_{k}= \left\lbrace  p,/ x_{p}^{k} \geq {dv}_p, \forall p=1,\ldots,K \right\rbrace$ \\
		    $Ed_{l}= \left\lbrace  p,/ x_{p}^{l} \geq {dv}_p, \forall p=1,\ldots,K \right\rbrace$\\
		     \CommentSty{ /* Apply Eq. (\ref{coenergydissip})*/} \\
		    $\text{CED}_{kl}=\frac{|\left(Ed_{k}\cap Ed_{l}\right) |}{|DV|}$
		}  
    \CommentSty{/*Apply threshold to decide to put a link, Eq. (\ref{linkactor})*/}\\
     \If {$\text{CED}_{kl} \geq \alpha$} 
     {
           $addLink(V, k, l)$
     }
    \Return \textbf{\textit{V}}
    }
}
 \caption{\textit{Sub-Community Discovering} }
 \label{transient}
 \end{algorithm}
 
\section{Using possibility theory to identify pivot(s)}
\label{possibility}
Our pivot identification methodology relies mainly in the feature of energy we defined for the actors/nodes. Pivots refer to nodes with "high" energy. These nodes have the opportunity to control the dynamic evolution of the network. If the energy of a pivot decreases or increases, the structure of the network might evolve towards a new direction allowing to suppress or add links between nodes. A simple question arises from the perspective of pivots identification: how much energy is necessary for a node to be classified as a pivot? We believe that it is very difficult to answer this question by analyzing only the amount of energy of each node. If someone would like to do so, he/she should build a  kind of threshold and apply the decision to put the label "pivot" on a node if its energy exceeds this limit.  This methodology weakens the objectivity of pivots' identification. To surround this difficulty, we propose a more robust identification scheme based on a link between probability theory and possibility theory.

The energy property can be viewed as a continuous random variable $X: \Omega \rightarrow V$, where $\Omega$ is the set of actors/nodes and $V$  the measurable function giving the real value of the energy as defined in equation \ref{energyactor}. $X$ does not return a probability. But we want to know, if it were the case, could this probability help achieving our goal. 

Probability theory is a valuable quantitative  tool to study randomness/uncertainty in random phenomena.
In this area,  we can find the probability of occurrence of different possible outcomes in an experiment. If a random variable returns a probability $P$, then $V=[0, 1]$ and $\displaystyle \sum_u P(X=u)=1$. As we defined the energy in equation \ref{energyactor}, the energy $E_i$ of actor $i$ is always between 0 and 1. So, we can take $V=[0, 1]$, but we do not have $\displaystyle \sum_{i}^{K} E_{i}=1$ ($K$ is the number of nodes). 

\subsection{Defining probability to characterize energy}
For the purpose to return a probability from $X$, we have just to normalize the energy by the size of $\mathbb{E}$, where $\mathbb{E}=\left\lbrace E_i, i=1\ldots,|\Omega|\right\rbrace$. At this point, the random variable $X$ returns a probability since the equation  $\displaystyle \sum_u P(X=u)=1$ holds.  We build this probability to characterize each energy of a node with a probability of occurrence in $[0, 1]$. Now, using this probability definition, our scope is to build a test to identify the potential pivots under each sub-community. Generally, a typical test consists of applying a threshold on the outcomes (probabilities of energy) to decide to put the label pivot on a node if the probability distribution takes a value higher than this threshold. The task to build this threshold is not straightforward even if,   for given nodes if the probability of their respective energy is 0.95, 0.6 and 0.3 for example, how could we decide to label a node as pivot, based only on the not obvious notion of "high" probability. Is 0.6 a "high" or "low" probability? By pointing out this example, we simply want to show that it is not evident to know the "best" value of the probability threshold in order to conclude if the node is a pivot or not. Nevertheless, one could, for simplicity, build an heuristic decision process where the threshold is set manually. With probability, the only evident decision we can take is, when the probability of a node is $P(X=u)=1$.
If this case happens, it means that there's only one pivot in the entire sub-community, since the other nodes have probability 0 and so this sub-community is built with only one node. Finally, we believe that pivots might be nodes with any probability, different to zero, "sufficient" to become member of a sub-community and to have the potential to change the dynamic evolution of the network. To go towards the direction of extracting the "best" level of probability measure, we think of an alternative related the area of possibility theory, which gives as another tool to represent uncertainty in a qualitative fashion. This tool can also helps to learn more about  the \textbf{incompleteness} that reflects the lack of information. We have seen above that, our defined probability doesn't give us the information about the limit to apply to detect pivots. So, the idea behind the use of this new scheme  is to associate to each node, in accordance to its energy, a degree of possibility which quantifies the level of "importance" of that node among the others.
\subsection{Detecting pivot by possibility degree}
Between probability and possibility, we can state a consistent principle is this terms: "\textit{what is probable should be possible}" \cite{zadeh}. 
This requirement can then be translated as follow:
\begin{equation}
\label{zadeh1}
P(A)\leq \Pi (A)  \qquad \forall A \subseteq \Omega
\end{equation}
where $P$ and $\Pi$ are, respectively, the probability and the possibility measure on the domain $\Omega$. In this case, $\Pi$ is said to \textbf{dominate} $P$. With $\Pi$, given nodes should have the maximum possibility degree, i.e. the value 1. In possibility theory the equality $\displaystyle \sum_u \Pi(X=u)=1$ is not guaranteed.
So, for any node, when its possibility degree reaches the maximum, we can robustly say that this node is a pivot since it is entirely sure that it might exist in the network. So, we see that the notion of "high" probability can be defined properly, since it corresponds to the maximum degree of possibility. 

A possibility measure, \cite{dubois}, $\Pi$ on $V$ is characterized by a possibility distribution $\pi$ : $V \rightarrow [0, 1]$, and is defined by:
\begin{equation}
\label{eq:pos}
  \forall A \subseteq V, \Pi (A) = sup \{ \pi(v), v \in A \}.
\end{equation}
If $V$ is a finite set, thus : $\forall A \subseteq V, \Pi (A) = max \{ \pi(v), v \in A \}$.
 The key concept of a possibility distribution is the preference ordering it establishes on $V$.
 Basically, $ \pi$ designates what one knows about the value of a given variable $X$, and
$\pi (v) > \pi (v')$ states that $X = v$ is more plausible than $X = v'$. 
When $\pi (v) = 0$, thus, $v$ is an impossible value of the variable $X$ while $\pi (v) = 1$ means that $v$ is one of the most plausible values of $X$. For us, identifying pivots is just a process to searching at these most plausible values.

Transforming a probability measure into a possibilistic one then amounts to 
choosing a possibility measure in the set $\Im(P)$ of possibility measures dominating $P$.
This should be done by adding a strong order preservation constraint, 
which ensures the preservation of the shape of the distribution:
\begin{equation}
\label{zadeh2}
p_i < p_j \Leftrightarrow \pi_i<\pi_j  \qquad  \forall i,j \in \lbrace1,\ldots,q\rbrace,
\end{equation}
where
 $p_i={P(\lbrace E_i} \rbrace)$ and $\pi_i={\Pi(\lbrace E_i} \rbrace),  \forall i \in \lbrace1,\ldots,K\rbrace$. It is possible to search for the most specific possibility distribution verifying (\ref{zadeh1}) and (\ref{zadeh2}).
 The solution of this problem exists, is unique and can be described as follows.  One can define a strict partial order $\textsf{P}$ on $\Omega$ represented by a set of compatible linear extensions $\Lambda(\textsf{P})=\lbrace l_u,  u=1,L\rbrace$. To each possible linear order $l_u$ , one can associate a permutation $\sigma_u$ of the set $\lbrace 1,\ldots,q \rbrace$ such that:
\begin{equation}
\label{zadeh3}
\sigma_u(i)< \sigma_u(j) \Leftrightarrow (\omega_{\sigma_{u}(i)},\omega_{\sigma_{u}(j)}) \in l_u,
\end{equation}
The most specific possibility distribution, compatible with the probability distribution $(p_1, p_2,\ldots, p_K)$ can then be obtained by taking the maximum over all possible permutations:
\begin{equation}
\label{zadeh4}
\pi_i=\displaystyle{\max_{u=1,L}}\displaystyle{\sum_{\lbrace j |\sigma_{u}^{-1}(j)\leq \sigma_{u}^{-1}(i)\rbrace}}p_j
\end{equation}
 Finally, the vector $(\pi_1, \pi_2,\ldots, \pi_K)$ gives us all the possibility degrees for the K nodes, corresponding to the $(p_1, p_2,\ldots, p_K)$ vector of probability of their energies. And pivots are nodes for which the possibility degree exceeds a given rate $\delta$. For this study we set this threshold $\delta=1$, it corresponds to the maximum value a possibility degree might be set. A more flexible and non-heuristic method might be to set $\delta$ to a value less than the maximum. But, here, we set $\delta$ to $1$ in order to show that in all situations or scenario, a pivot must be found.
 
For a thorough view of possibility theory, we recommend the reader to \cite{zadeh,dubois,zadeh2}.

\section{Validation}
\label{validation}
We validate our approach on the  real world collection of data coming from \textbf{Reddit.com}  \cite{sitis15}. We use several samples of different sizes and, build four scenarios A, B, C and D with dimension ($N\times K$, $N$ the number of actors and $K$ the number of events) $10\times 15$, $10\times 150$, $10\times 500$ and $10\times 1200$ respectively. In Table .\ref{tableEventInitial}, we give an idea on the content of the data, in each column vector, we have the total amount of submissions to an image by the set of actors.
 \begin{table*}
    \centering
\renewcommand{\arraystretch}{0.7}
\begin{tabular}{|c|c|c|c|c|c|c|c|c|c|c|c|c|c|c|c|}
\hline
  \multicolumn{1}{|c|}{\backslashbox{Actors}{\vrule width 0pt height 1.25em events}}& $e_1$  & $e_2$ & $e_3$  & $e_4$ & $e_5$ & $e_6$ & $e_7$ & $e_8$ & $e_9$ & $e_{10}$&$e_{11}$&$e_{12}$&$e_{13}$&$e_{14}$&$e_{15}$\\
   \hline
    1 &11     &0&    11  &   4 &    0&     2   &  0 &    4 &   18 &    2 &    0  &   6 &   16 &    1 &  0\\
    2  &5     &0   &  0  &   0  &   0   &  0 &    1  &   0   &  0 &    0 &    2  &   0 &    2   &  0  & 0\\
    3  &1     &0   &  2 &    0 &    3  &   1  &   1  &   0  &   1    & 0  &   0  &   2  &   1  &   0 &  2\\
    4  &4     &1  &   0   &  0  &   0 &    1  &   2 &    0   &  7  &   0   &  1  &   0   &  9 &    1&   0\\
    5  &1     &0   &  0   &  0  &   0  &   0 &    0  &   0  &   0 &    0  &   0  &   0  &   0  &   0&   0\\
    6  &1     &0    & 0  &   0  &   0  &   0 &    0  &   0  &   0 &    0 &    0  &   0 &    0  &   1 &  0\\
    7  &0     &2   &  0   &  0   &  1  &   0 &    0   &  0  &   0  &   0 &    0  &   4 &    0  &   0 &  0\\
    8  &0     &0   &  0    & 0   &  0  &   0 &    0   &  0  &   0   &  0 &    0  &   0  &   0   &  0 &  0\\
    9  &0     &0   &  0   &  0   &  0   &  0 &    0  &   0   &  0 &    0  &   0 &    0  &   0  &   0  & 0\\
    10 &0     &0 &    0 &    0 &    0  &   0 &    0 &    0 &    0  &   0  &   0  &   0 &    0  &   0 &  0\\
   \hline
\end{tabular}
\caption{Activities and amount of actor participation to submissions on events. Scenario A.}
\label{tableEventInitial}
\end{table*}

Two levels of information can be retrieved from the results. In the first level, we have the results about the formation of the underlying sub-communities. This result corresponds to the natural clustering of the different nodes according to the energy provided by each of them. The second level of information refers to the characteristics of links and nodes inside the given sub-groups. This refinement provides useful information when one wants to emphasize and explore   some parts of the network.
\subsection{Information about the formation of sub-communities}
One of the main objectives of social network analysis is related to clustering in order to study the similarities inside the network \cite{mcgloin2010overview,wasserman1994social}. So, the first result is about the formation of sub-communities. The graphs in Fig. \ref{impactcoenergydissip} show all the groups we discover with the different scenarios. In Fig. \ref{in15}, we see a sparse sub-community within two dense sub-communities. The term 'sparse' refers to a group of nodes with different levels of energy. When we observe a group of linked nodes with the same value for their energy, we consider this group as an inner dense sub-community; here, we have $\{1,3,7\}$ and $\{2,3,7\}$. 

\begin{figure*}
 \begin{center}
  \subfigure[{Co-Energy between actor 1 and actor 2}\label{co12}]{\includegraphics[scale=0.25152]{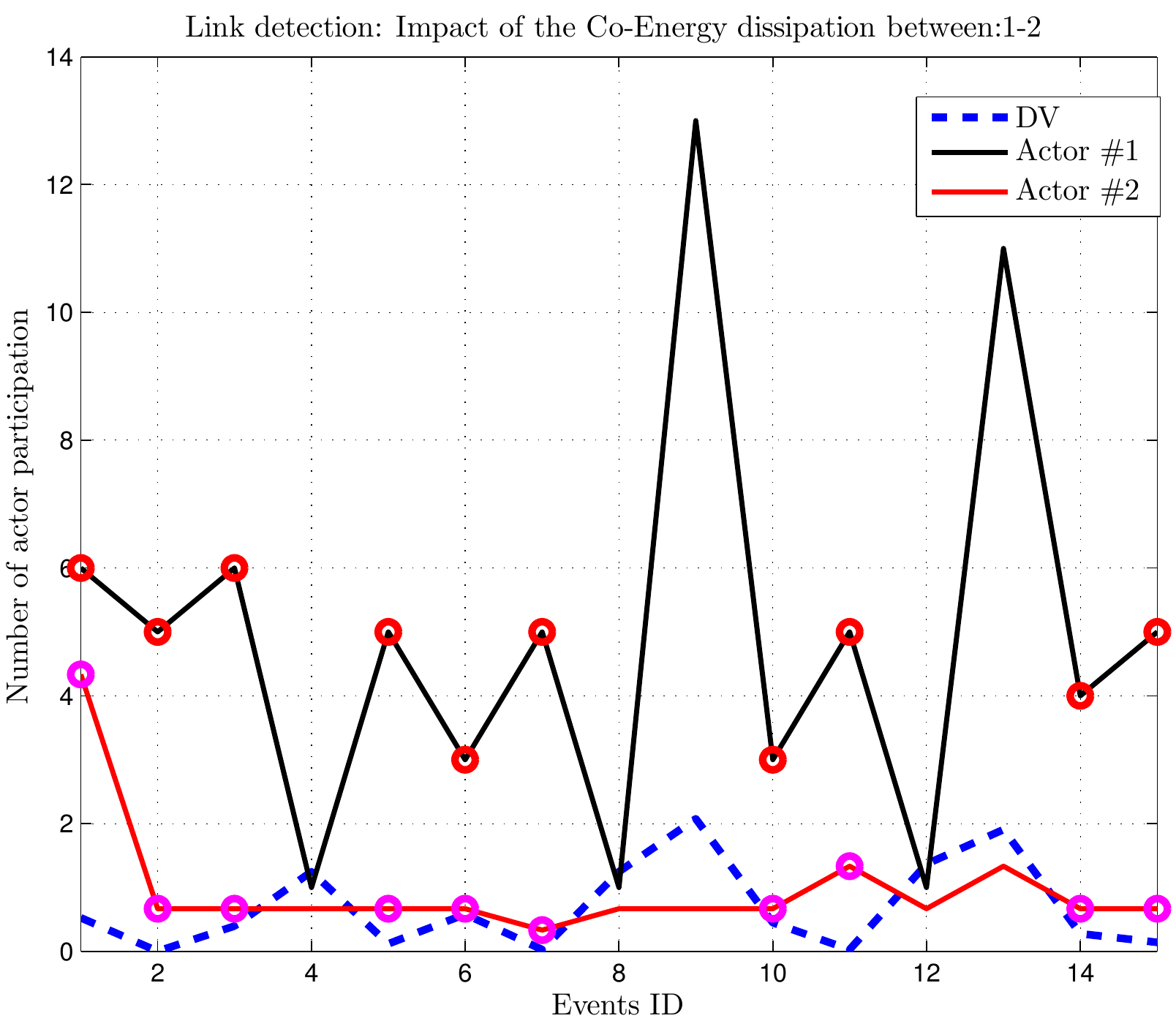}}
  \subfigure[{Co-Energy between actor 1 and actor 3}\label{co13}]{\includegraphics[scale=0.25152]{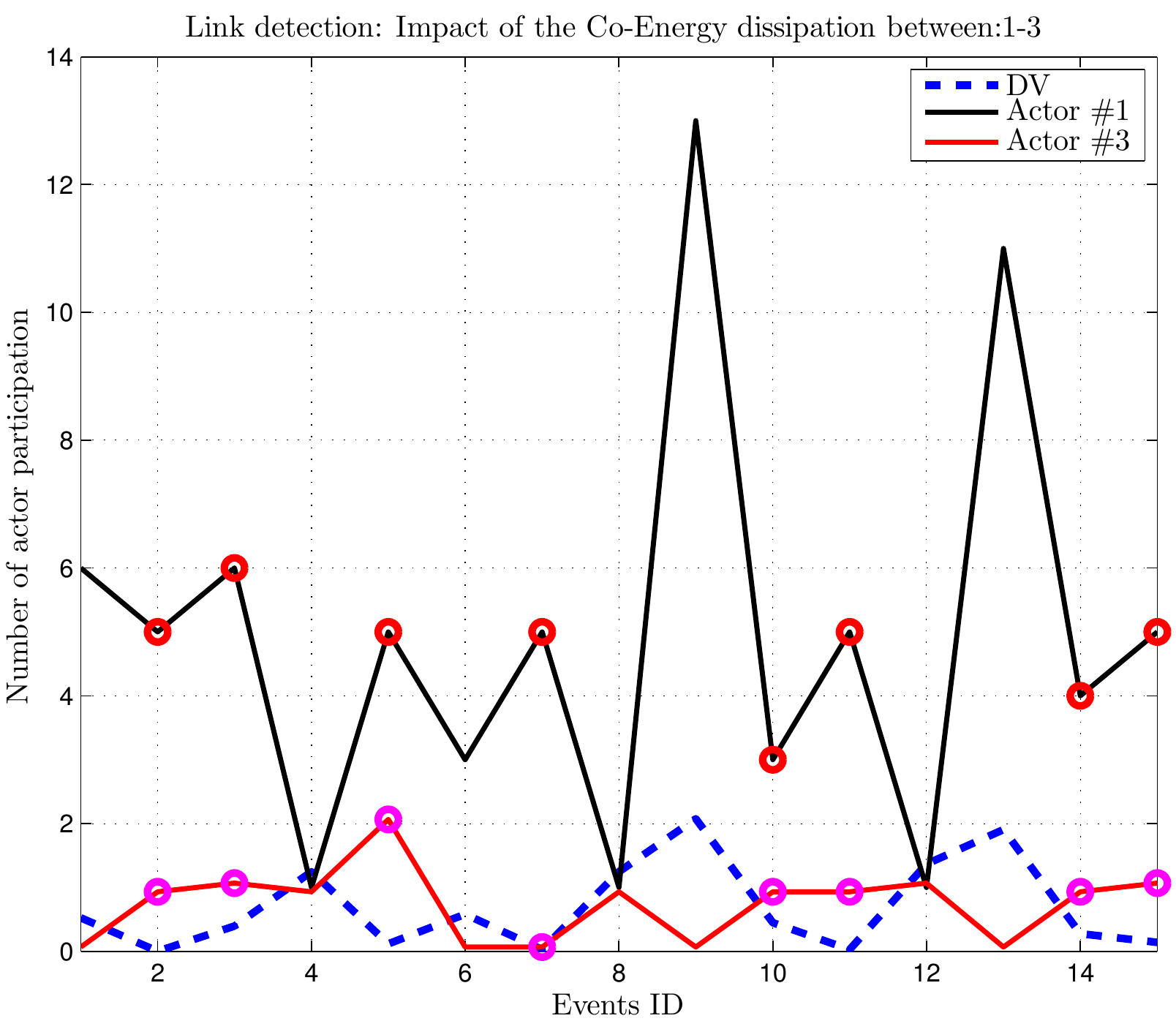}}
  \subfigure[{Co-Energy between actor 2 and actor 7}\label{co27}]{\includegraphics[scale=0.25152]{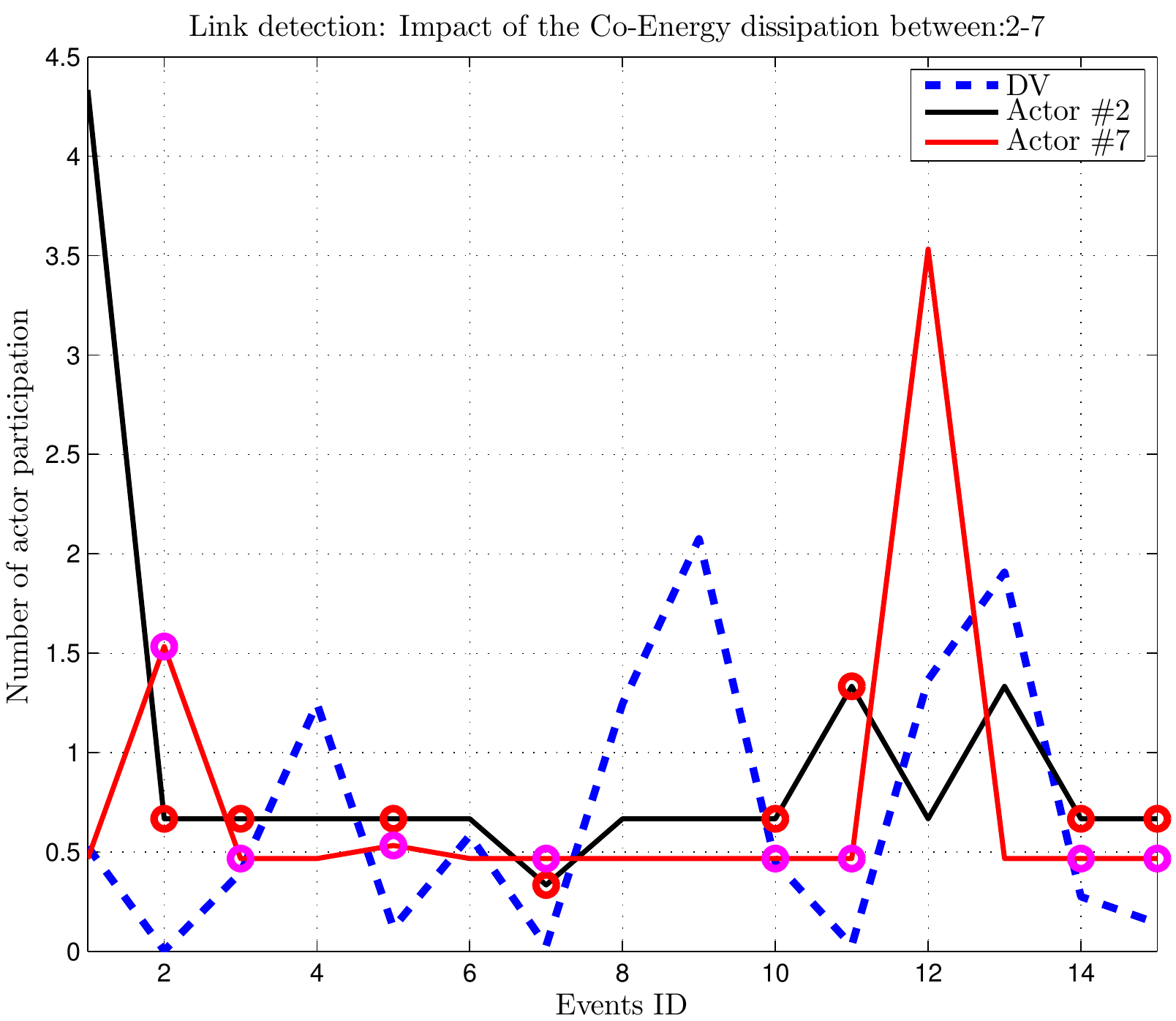}}
  \subfigure[{Co-Energy between actor 4 and actor 7}\label{co47}]{\includegraphics[scale=0.25152]{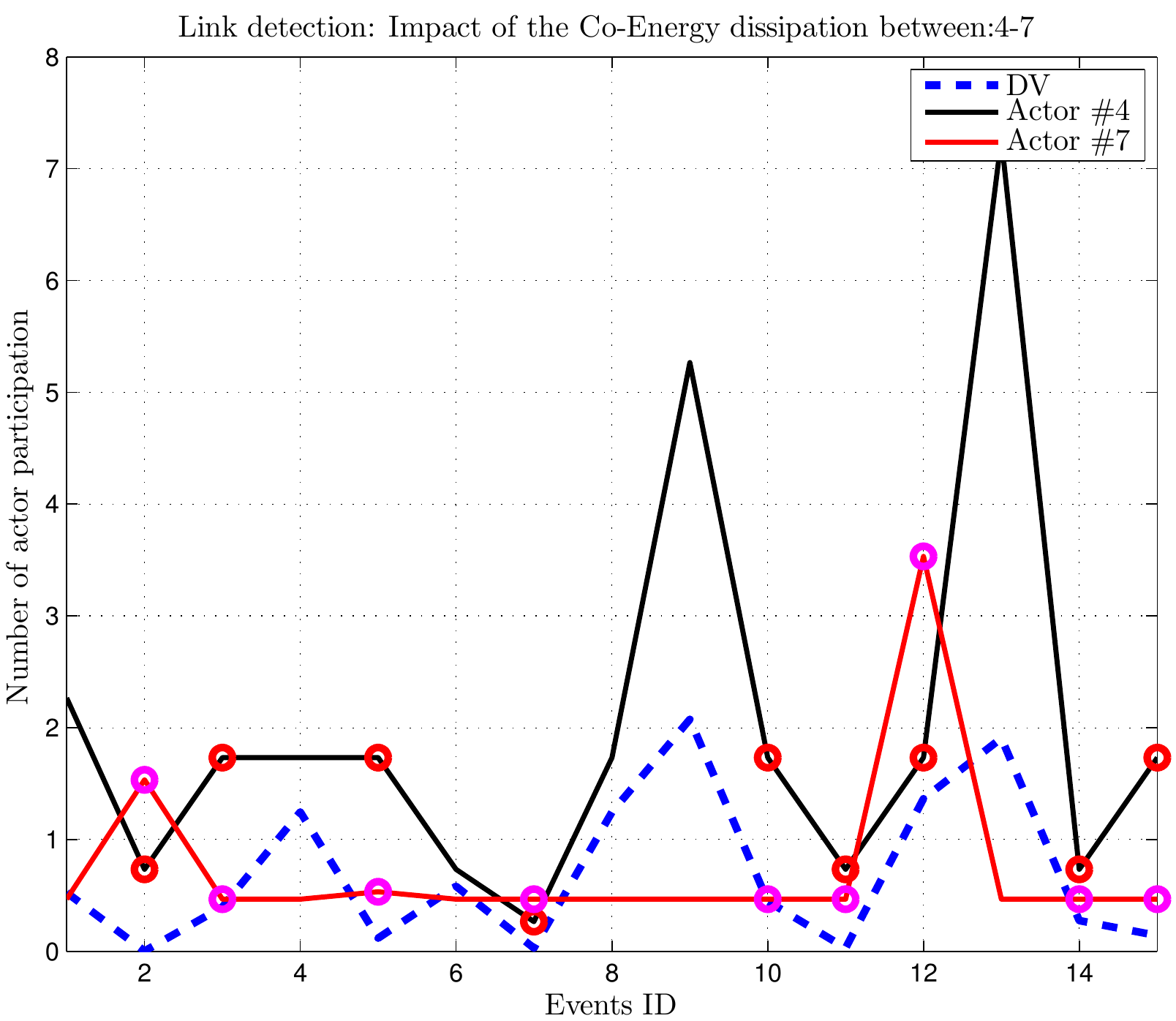}}
 \caption{Impact of the Co-Energy dissipation between two actors in the link detection phase.
 The circles represents events where both the two actors have their co-energy higher than the reference value in the decision variable DV. These actors are linked since their co-energy dissipation concern more than half of the events.}
\label{impactcoenergydissiponevents}
 \end{center}
\end{figure*}

In graphs of Fig. \ref{impactcoenergydissiponevents}, we draw the co-energy participation between two nodes to emphasise the fact the role of the Decision Variable DV have to set a potential link. Considering all the events at the same time, a link can be put between two nodes if the amount of participations of both two nodes, for the same set of events, exceeds by far the reference point given by the decision variable. We use  circles to identify the events where the energy of the actors is higher than the reference point. Clearly, for most of the given events, if the energy of each node exceeds the value of the DV for that events, then we put a link between the two nodes.
\subsection{Information about the intrinsic behavior inside sub-communities}
\begin{figure*}
 \begin{center}
  \subfigure[{A sparse sub-community within two dense intra sub-communities. Scenario A with 15 events.}\label{in15}]{\includegraphics[scale=0.152]{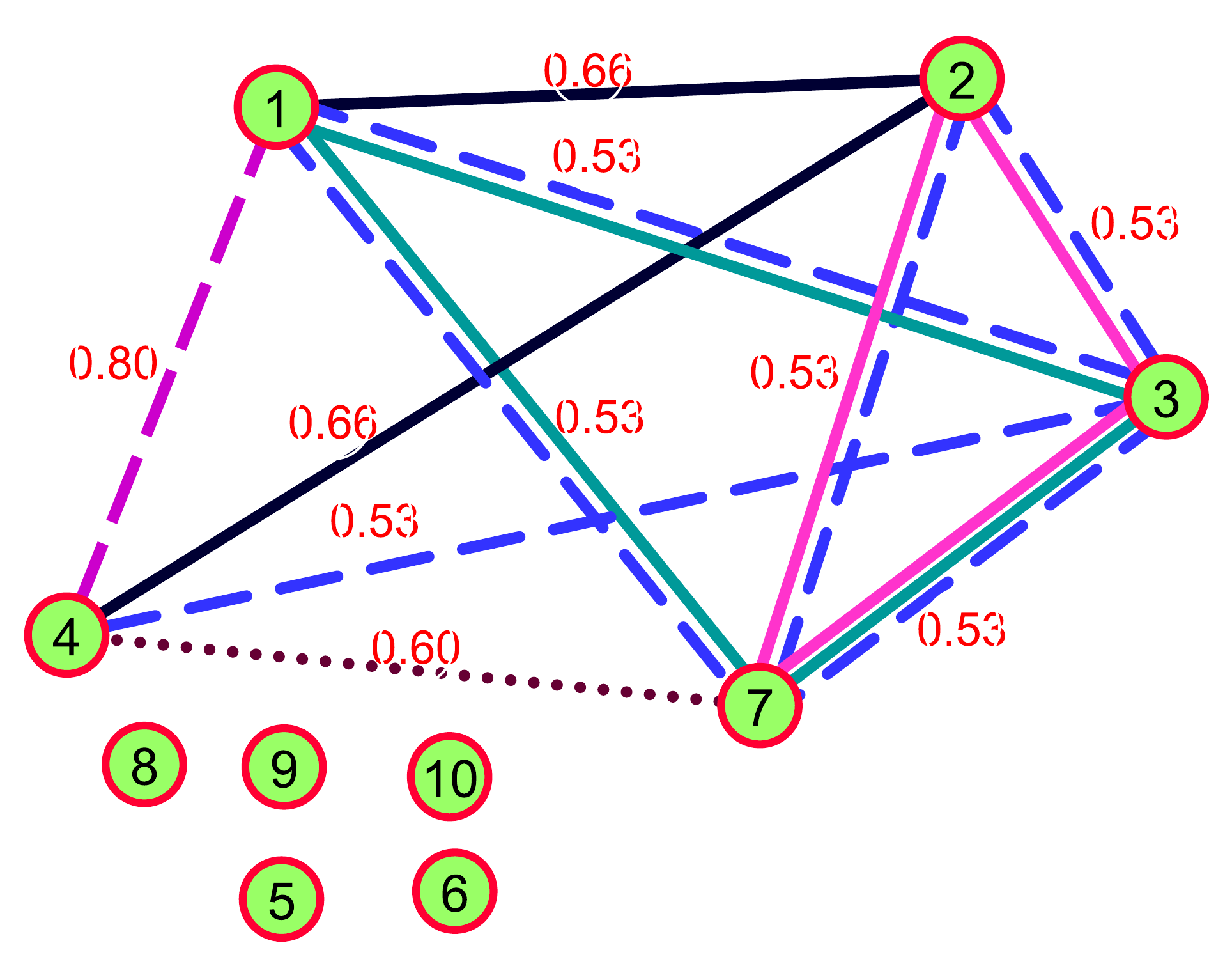}}
  \subfigure[{A sparse sub-community. Scenario B with 150 events.}\label{in150}]{\includegraphics[scale=0.152]{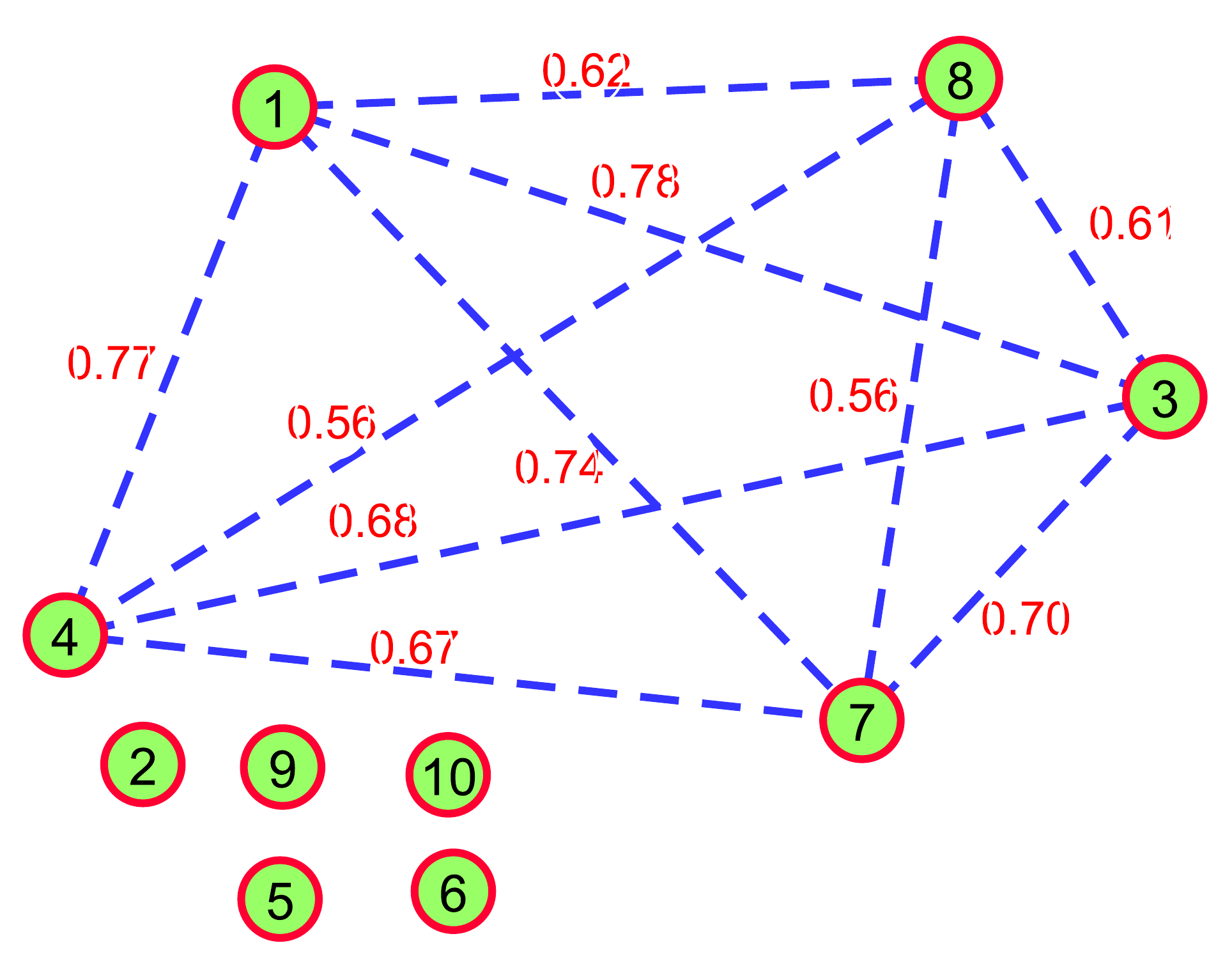}}
  \subfigure[{A sparse sub-community. Scenario C with 500 events.}\label{in500}]{\includegraphics[scale=0.152]{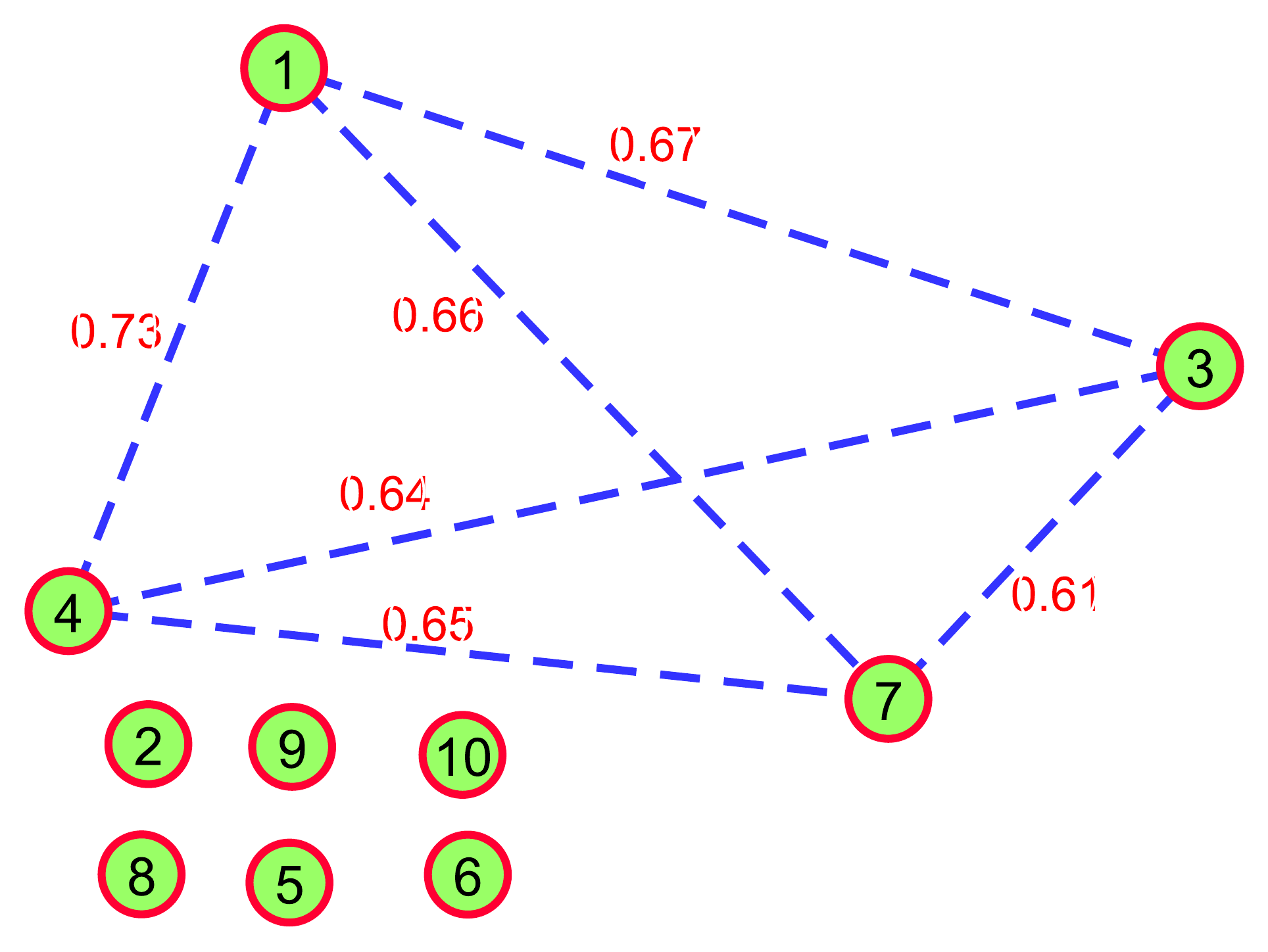}}
   \subfigure[{A sparse sub-community. Scenario D with 1200 events.}\label{in1200}]{\includegraphics[scale=0.152]{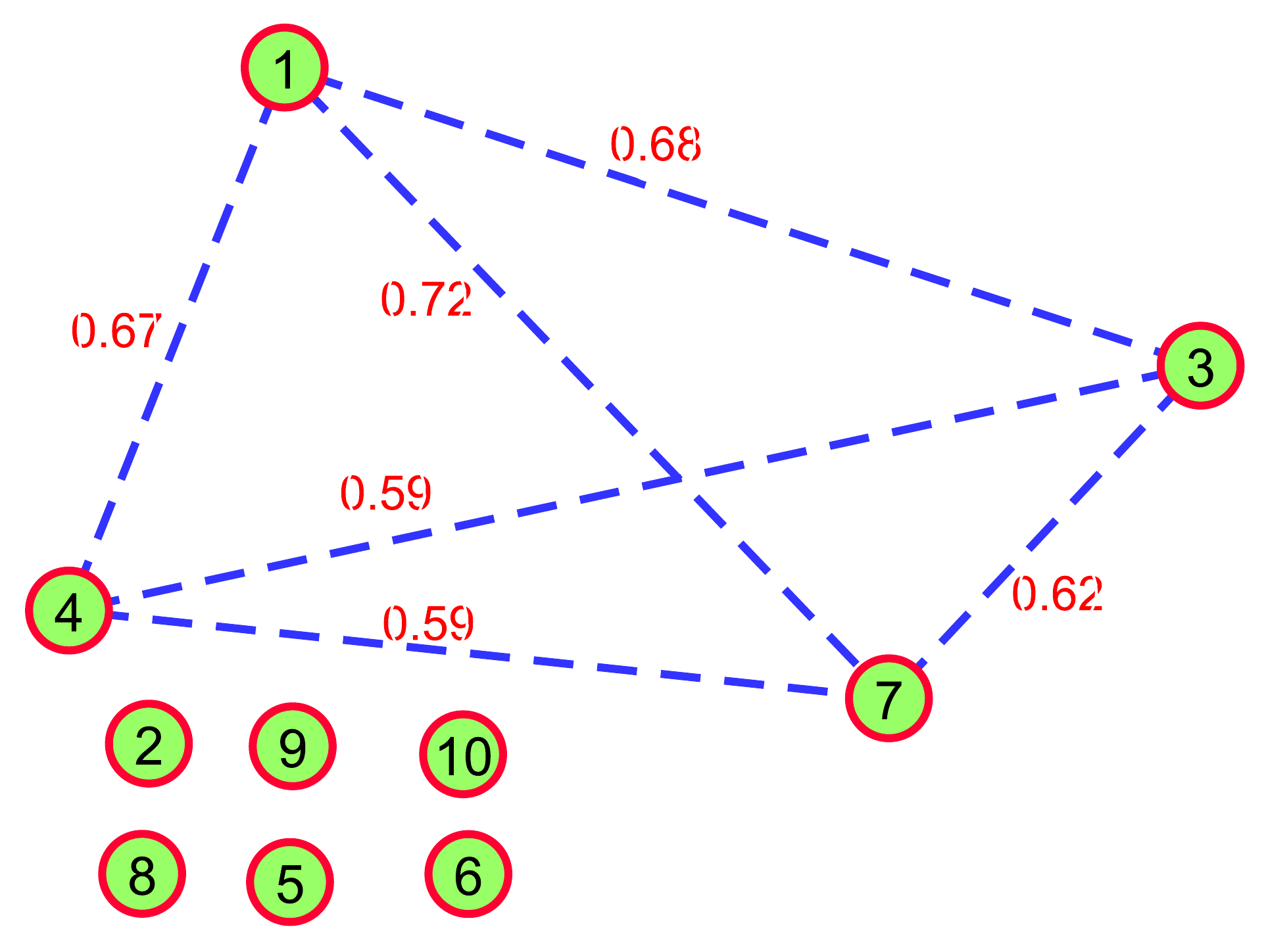}}
 \caption{Results of the sub-community members identification. In scenario A \ref{in15}, we can observe two dense sub-communities. In the other scenarios (B, C and D), we obtain only a sparse sub-community. The different scenarios are built with different size of the vector of events. }
\label{impactcoenergydissip}
 \end{center}
\end{figure*}

The second level of information this technique might deliver is about the dynamics of nodes and their relation inside the detected sub-groups. So, another result is related to the boundary of each detected link. In Table \ref{tabtotalenergy}, we have for each node, its total energy, i.e. the amount of power of this node to participate to all events. By observing carefully this table and the graphs of the sub-communities, we see that the bounds of a link is the interval $\left[a,b\right]$, where $a$ and $b$ are the respective total energy of the specified nodes. And so, the score/weight of a link is always inside this interval, as we can observe for all links detected. For example, in Fig. \ref{in1200}, the link between node \#4 (with energy 0.68) and node \#7 (with energy 0.74)has a weight of 0.59 and its bounded interval is $\left[0.68,0.75\right]$. As the network evolves, whenever the energy of a node belongs out of the interval, the link will disappear. By inspecting frequently the evolution of the bounded interval, one can retrieve useful information about the degree of importance of the different nodes by analyzing their energy.

Each of the other scenarios (B, C and D) give also a sparse sub-community. 

\begin{table*}[t]
\begin{center}
\caption{Total Energy of each actor for the different scenarios.}
\label{tabtotalenergy} 
\begin{tabular}{|c|c|c|c|c|c|c|c|c|c|c|c|}
 \hline
 \multicolumn{10}{c}{Energy of the different actors (Eq. \ref{energyactor})} \\
 \hline
 & $a_1$ & $a_2$& $a_3$  & $a_4$& $a_5$ & $a_6$ & $a_7$ & $a_8$ & $a_9$ & $a_{10}$ &\text{Number of events}      \\
 \hline
 $\text{E}_{i}$& 0.80 & 0.66   & 0.53& 1& 0.26& 0.40&0.60&0&0&0&15 \\
 $\text{E}_{i}$& 0.88 & 0.25   & 0.86& 0.77& 0.46& 0.25&0.74&0.62&0.15&0.13&150 \\
  $\text{E}_{i}$& 0.79 & 0.26   & 0.79& 0.74& 0.23& 0.19&0.69&0.46&0.11&0.06&500 \\
 $\text{E}_{i}$& 0.86 & 0.27   & 0.70& 0.68& 0.25& 0.23&0.74&0.31&0.97&0.55&1200 \\
 \hline
 \end{tabular}
\end{center}
\end{table*}

\subsection{Results and discussion about the pivots}
We apply the detection process to identify pivots to the data samples for scenarios A, B, C, D, E, F and G. We have arbitrarily chosen the  size of the sample and the interval where data come from. We just want to show cases where there's one pivot or more.   The table \ref{tabscenarios} resumes the data selection scheme, and in we put the results of the detection procedure in table \ref{tabpivot} where the variable NBE refers to the total amount of participation on events by nodes, $p$ is the probability of the energy and $\pi$ the corresponding degree of possibility. The results give many useful information. For  the scenarios A, B, C and D, we extract only one pivot which have the particularity to be the node with the highest amount of activities in the network. The pivot also has the highest probability of energy but, the value of this probability seems to be "very low" in a pure point-of-view of the probability theory. For example, for scenario A, the pivot has a probability $p=0.30$ but its possibility degree reaches $\pi=1$. For scenarios A, B, C and D, we find only one pivot and it corresponds to the node with the highest probability and which have perform the greater amount of activities in the network. But this findings are not a generality since we see, in the scenario G, the pivot has the highest probability but it does not have the greatest amount of activities. This discover  says clearly that, learning only the amount of activities is not a sufficient process to analyze a network in order to detect links or to put a level of importance/quality to nodes. In scenarios E and F, we detect twos pivots which have not the same amount of actions but they have the same probability of energy. We think that this situation is due to the fact that the quality of node depends not only to its level of participation on events but, it depends also to its interaction with other nodes. So one must have two (or more) nodes with different level of actions on events but they have the same behavior regarding to other nodes.
 
\begin{table}
\footnotesize{
\begin{tabular}{|c|c|c|c|c|c|c|c|}
  \hline 
  Scenario     & A      &  B     &     C  &  D     &  E     &   F & G \\
  \hline
  Size       & 15 & 150 & 500 & 1200 &155  & 255& 555\\
  $interval$ &[1:15]& [1:150] & [1:500] & [1:1200] & [1100:1254] &[1000:1254]  &[700:1254] \\
  \hline
\end{tabular}
}
\caption{Information about the selection of the different data samples }
\label{tabscenarios}
\end{table}

 \begin{table*}[t]
\begin{center}
\caption{Detection of pivots when degree of possibility $\pi=1$}
\label{tabpivot} 
\begin{tabular}{|c|c|c|c|c|c|c|c|c|c|c|c|}
 \hline
Nodes $i$ & $a_1$ & $a_2$& $a_3$  & $a_4$& $a_5$ & $a_6$ & $a_7$ & $a_8$ & $a_9$ & $a_{10}$ &\text{scenario}      \\
 \hline
  $NBE$ & 75   & 10     & 14   & 26   & 1  & 2   & 7    &0  &0  &0  &\multirow{4}{0.4cm}{A} \\
  $p$   & 0.30 & 0.10   & 0.25 & 0.25 & 0  & 0   & 0.10 &0  &0  &0 & \\
  $\pi$ & 1    & 0.20   & 0.70 & 0.70 & 0  & 0   & 0.20 &0  &0  &0 & \\
  \hline
  pivot & Yes& No& No& No& No& No&No&No&No&No\\
 \hline
   $NBE$ & 591   & 28     & 218   & 142   & 44  & 31   & 117    &53  &10  &8  &\multirow{4}{0.4cm}{B} \\
  $p$   & 0.42 & 0.06   & 0.18 & 0.23 & 0  & 0   & 0.11 &0  &0  &0 & \\
  $\pi$ & 1    & 0.06   & 0.35 & 0.59 & 0  & 0   & 0.17 &0  &0  &0 & \\
 \hline
   pivot & Yes& No& No& No& No& No&No&No&No&No\\
 \hline
  $NBE$ & 1638   & 108     & 668   & 348   & 120  & 63   & 348    &170  &27  &16  &\multirow{4}{0.4cm}{C} \\
  $p$   & 0.39 & 0.11   & 0.17 & 0.22 & 0  & 0   & 0.11 &0  &0  &0 & \\
  $\pi$ & 1    & 0.22   & 0.38 & 0.61 & 0  & 0   & 0.22 &0  &0  &0 & \\
  \hline
    pivot & Yes& No& No& No& No& No&No&No&No&No\\
 \hline
   $NBE$ & 3659   & 257     & 1596   & 773   & 206  & 164   & 890   &425  &59  &40  &\multirow{4}{0.4cm}{D} \\
  $p$   & 0.39 & 0.11   & 0.17 & 0.22 & 0  & 0   & 0.11 &0  &0  &0 & \\
  $\pi$ & 1    & 0.22   & 0.38 & 0.61 & 0  & 0   & 0.22 &0  &0  &0 & \\
  \hline
    pivot & Yes& No& No& No& No& No&No&No&No&No\\
 \hline
 \hline
   $NBE$ & 428   & 34     & 204   & 95   & 26  & 40   & 52    &40  &6  &4  &\multirow{4}{0.4cm}{E} \\
  $p$    & 0.38  & 0      & 0.38  & 0    & 0.06  & 0.06   & 0.06 &0.06  &0  &0 & \\
  $\pi$ & 1    & 0   & 1 & 0 & 0.25  & 0.25   & 0.25 &0.25  &0  &0 & \\
  \hline
    pivot & Yes& No& Yes& No& No& No&No&No&No&No\\
 \hline
   $NBE$ & 702   & 50     & 321   & 140   & 38  & 54   & 122    &82  &13  &8  &\multirow{4}{0.4cm}{F} \\
  $p$   & 0.32 & 0   & 0.32 & 0.07 & 0  & 0   & 0.07 &0.15  &0  &0.07 & \\
  $\pi$ & 1    & 0   & 1 & 0.23 & 0  & 0   & 0.23 &0.38  &0  &0.23 & \\
  \hline
    pivot & Yes& No& Yes& No& No& No&No&No&No&No\\
 \hline
   $NBE$ & 1569   & 127     & 663   & 335   & 72  & 81   & 405    &170  &22  &16  &\multirow{4}{0.4cm}{G} \\
  $p$   & 0.32 & 0   & 0.38 & 0 & 0.06  & 0   & 0.18 &0.06  &0  &0 & \\
  $\pi$ & 0.62    & 0   & 1 & 0 & 0.12  & 0   & 0.31 &0.12  &0  &0 & \\
  \hline
     pivot & No& No& Yes& No& No& No&No&No&No&No\\
 \hline
 \end{tabular}
\end{center}
\end{table*}

\section{Conclusion}
\label{conclusion}
In this work, we have extent a new technique related to an extended version of principal component analysis to build a methodology for the purpose of community detection in a social network. The initial work built a technique more elaborated to run within stochastic process than the classical PCA which is designed originally to solve the problem of dimensionality reduction for univariate dataset. The main innovation of the work is manifold: (i) we define the notion of co-energy between two nodes to quantify the intensity of their relation, (ii) we can also extract the proper energy of a given node to know how it influences the overall community, (iii) technically, the KL-PCA technique makes possible to build a decision variable and to form a state model from which we apply a decision process to identify each link. The introduction of the notion of energy make possible to see potential intra sub-communities (i.e. nodes with the same co-energy) inside a sub-community; (iv) each detected link is bounded, so we know how much energy is necessary to maintain a link over time. In this complementary study, we show that a possibility distribution can be properly defined from the energy to solve an interesting feature i.e, the problem of identifying pivots which have the main impact in the dynamic nature of the network. From this work, we plane to learn the impact of the number of pivots in a given sub-community and between sub-communities to face the idea related to their impact on a network distributed in many geographic area.


\bibliographystyle{bmc-mathphys} 
\bibliography{references_2}
\end{document}